\documentclass[onecolumn,superscriptaddress,aps,pra,showpacs,floatfix]{revtex4}

\usepackage{latexsym}
\usepackage{color}
\usepackage{graphicx}
\usepackage{cancel}
\usepackage{bm}
\usepackage{maybemath}

\newcommand{\ee}{\mathrm{e}}
\newcommand{\ii}{\mathrm{i}}

\newcommand{\calC}{\mathcal{C}}

\newcommand{\calL}{\mathcal{L}}

\newcommand{\plus}{{\mbox{{\bf{\tiny +}}}}}

\begin{document}

\title{Neutrino Helicity Reversal and Fundamental Symmetries}

\author{U. D. Jentschura}

\affiliation{Department of Physics,
Missouri University of Science and Technology,
Rolla, Missouri 65409, USA}

\affiliation{MTA--DE Particle Physics Research Group,
P.O.Box 51, H-4001 Debrecen, Hungary}

\author{B. J. Wundt}

\affiliation{Department of Physics,
Missouri University of Science and Technology,
Rolla, Missouri 65409, USA}

\begin{abstract}
A rather elusive helicity reversal occurs
in a gedanken experiment in which a massive left-handed Dirac neutrino,
traveling at a velocity $u < c$, is overtaken on a highway by a speeding
vehicle (traveling at velocity $v$ with $u < v < c$).
Namely, after passing the neutrino, looking back, one would see a
right-handed neutrino (which has never been observed in nature).  The
Lorentz-invariant mass of the right-handed neutrino is still the same as before
the passing. The gedanken experiment thus implies the existence of
right-handed, light neutrinos, which are not completely sterile.
Furthermore, overtaking a bunch of massive right-handed Dirac neutrinos
leads to gradual de-sterilization.
We discuss the helicity reversal and the
concomitant sterilization and de-sterilization
mechanisms by way of an illustrative example
calculation, with a special emphasis on massive
Dirac and Majorana neutrinos. We contrast the formalism with a modified Dirac neutrino
described by a Dirac equation with a pseudoscalar mass term
proportional to the fifth current.
\end{abstract}


\maketitle


%
%
\section{Introduction}

The neutrino (here collectively used in order to denote a mass eigenstate of
the neutrino field) is the only particle in the (extended) Standard Model for
which an appropriate first-quantized description has not yet been completely
determined, and the observation of nonvanishing neutrino masses has not
simplified the task (for pertinent overview, see 
Refs.~\cite{KaPeGB1989,GiKi2007,BaMaWh2012,Zu2012}). 
Possible candidate equations include (i)~the Dirac
equation with a standard, scalar mass term~\cite{Ro2012}, (ii)~the Majorana
equation (which would imply that neutrinos are equal to their own antiparticle,
see Ref.~\cite{Pa2011ajp}), and (iii)~generalized Dirac equations with a
pseudoscalar mass term~\cite{ChHaKo1985,ChKoPoGa1992,Ko1993,ChKo1994,%
JeWu2012epjc,JeWu2012jpa,JeWu2013isrn}.  The latter equation leads to a
manifest helicity dependence of the anti-commutators of the field operators and
to a superluminal (tachyonic) dispersion relation, which has recently been
discussed in Ref.~\cite{JeWu2013isrn}.  Recent claims regarding a possibly
superluminal nature of the neutrino, which eventually turned out to be in
error, have inspired theorists to revisit theoretical ideas whose relevance
extends beyond relevance to the experimental claims.  However, in order to keep
things in perspective, we should remember that the origin of these ideas dates
back about three decades~\cite{ChHaKo1985}.  Finally, (iv)~further modified
Dirac equations with small Lorentz-breaking terms have also been discussed in
the literature~\cite{KoMe2012}.

In the current article, our focus will be on a comparison of the helicity
suppression in models~(i) and~(iii), while the Majorana character of a
neutrino~(ii) is commonly associated to the seesaw
mechanism~\cite{PaSa1974,SeMo1975,MoSe1979,BiGi2012}, which generates neutrino
masses and the mass gap of left-handed and right-handed neutrinos by
spontaneous symmetry breaking.  It leads to an effective suppression of the
right-handed neutrino interaction, in view of a large mass separation of the
different helicities.  Namely, right-handed neutrinos acquire a mass of order
$\Lambda_{\rm GUT}$, where $\Lambda_{\rm GUT}$ is the grand unification scale,
while left-handed neutrinos acquire a mass of the order of $v^2/\Lambda_{\rm
GUT}$, where $v$ is the Higgs vacuum expectations value.  From the point of
view of fundamental symmetries, a Majorana character of the neutrino, combined
with the seesaw mechanism, would solve two issues simultaneously.  Namely,
(i)~it would explain why the neutrino masses are so small as compared to the
masses of other particles in the Standard Model, and (ii)~it would offer a very
natural interpretation for the elusive helicity reversal of a neutrino
overtaken by a speeding vehicle on a highway. The right-handed state would
naturally be interpreted as a right-handed anti-neutrino, i.e., as a physical
state of the neutrino's own (identical) anti-particle.

However, the introduction of a Majorana neutrino into an extended Standard
Model is not as innocent as it seems.  In particular, the Majorana mass term
violates lepton number, which is a global symmetry that tracks the difference
between particles and antiparticles (an excellent overview is presented in
Ref.~\cite{Pa2011ajp}).  While there is nothing sacred about global symmetries,
``lepton number conservationalists'' might find the lack of a proper
distinction of particles and antiparticles disturbing. A Majorana neutrino
would be the only spin-$\tfrac12$ particle in an (extended) Standard Model
which is equal to its own antiparticle (the original standard model called for
a Weyl neutrino).  Furthermore, a Majorana neutrino is described by an equation
which, on the level of first quantization, does not allow plane-wave solutions
of the form $u \, \exp(-\ii k \cdot x)$, where $u$ is a spinor (or polarization
vector in the case of spin-$1$ particles) and $k \cdot x = E \, t - \vec k
\cdot \vec r$ is the scalar product of energy-momentum and space-time
coordinate (see also~\ref{appc}).  Recent measurements set rather strict bounds
for the magnitude of the Majorana mass
terms~\cite{Gi2012,AuEtAl2013,GaEtAl2013,MaEtAl2013} as inferred from
neutrinoless double beta decay experiments.

In order to put things into perspective, we recall that in the ``good old
times'', neutrinos were supposed to be massless Weyl fermions which come in
only one helicity state. Weyl fermions transform according to the fundamental
$(\tfrac12,0)$ representation of the Lorentz group. The 
well-known discovery of neutrino oscillations implies that
the neutrinos need to have a nonvanishing mass.  If neutrinos are Majorana
fermions, then the right-handed states could be interpreted as antineutrinos
because Majorana fermions are equal to their own antiparticles, but this
interpretation is unavailable if neutrinos are Dirac fermions.  Because of the
historical, proverbial association of very high speeds with the ``autobahn'',
we would like to refer to the underlying question as the ``helicity reversal
paradox'' or even the ``autobahn helicity paradox'' while stressing that the
problem has been raised here before in the scientific literature. 
Other possible designations, inspired by Ref.~\cite{LEPUS}, 
would include the ``rabbit paradox''
in light of the fact that high propagating velocities needed to 
overtake a neutrino are commonly associated with the physical 
abilities of a species known as the ``lepus'' or ``lepus curpaeums'' (in Latin).
Indeed, the elusive helicity reversal of a neutrino, overtaken on a highway by speeding
vehicle, has given rise to a few questions discussed by Goldhaber and
Goldhaber~\cite{GoGo2011}. In Ref.~\cite{Fe1998q76}
(labelled as ``question~\#76'' regarding ``neutrino mass and helicity'' 
in the American Journal of Physics), the author raises the question as to how the helicity
flip upon (hypothetically) overtaking a left-handed neutrino should be
interpreted physically: If only left-helicity neutrino states take part in the
weak interaction and helicity is not conserved upon a Lorentz boost, then is
the theory of the weak interaction self-consistent? 
Although this paradox has been around for as long as theorists have conceived
of massive neutrinos and has often been discussed and argued in qualitative
terms, it is somewhat surprising that we were unable to find a direct
quantitative treatment of the relevant physics in the academic literature.
We note that helicity (not
chirality!) remains a good quantum number for a massive neutrino.
An attempt to address the questions was presented in Ref.~\cite{Ho1998q76}.
In the helicity basis (see Chap.~23 of Ref.~\cite{BeLiPi1982vol4},
Sec.~2.8.1 on p.~28 of Ref.~\cite{GiKi2007} and Sec.~3 of 
Ref.~\cite{JeWu2013isrn}), plane-wave neutrino states are characterized by the
wave vector $\vec k$ and the helicity quantum number~$\sigma$.  The wave vector
$\vec k$ constitutes a quantum number in a continuous spectrum.  In general,
the energy $E \to E'$, the wave vector $\vec k \to \vec k'$ and also the
helicity $\sigma \to \sigma'$ change upon a Lorentz boost (transformation to a
different Lorentz frame). In particular, the helicity quantum number $\sigma$
reverses sign (``flips'') as one ``overtakes'' the neutrino, by going into a
reference frame which travels at a velocity $v > u$ in the direction of the
neutrino velocity $u < c$ (one has $u < v < c$).  A right-handed (helicity)
Dirac neutrino is not completely sterile because the projector $\gamma^\mu
(1-\gamma^5)$ in the weak-interaction Lagrangian projects onto left-handed
chirality, not helicity, states, and right-handed (helicity) massive subluminal
Dirac neutrinos are never in a perfect right-handed chirality eigenstate. 

Here, we start with a discussion of the helicity suppression for a standard
Dirac neutrino in Sec.~\ref{sec2}. The case of a Majorana neutrino is discussed
in Sec.~\ref{sec3}. The necessary formalism for tachyonic Dirac neutrinos is
recalled in Sec.~\ref{sec4}.  We show that real as opposed to virtual
spin-$1/2$ particles described by the tachyonic Dirac equation are always
left-handed.  Conclusions are given in Sec.~\ref{sec5}, while~\ref{appa} is
devoted to a discussion of the transformation into a rotating frame of
reference.  Further physical consequences of a neutrino theory based on the
generalized Dirac equation are discussed in~\ref{appb}.  Finally,~\ref{appc} is
devoted to the Majorana equation.  Units with $\hbar = c = \epsilon_0 = 1$ are
used throughout this article. 

%
%
\section{Dirac Equation and Helicity Suppression}
\label{sec2}

We start from the standard Dirac equation with a scalar mass term,
\begin{equation}
\left( \ii \gamma^\mu \partial_\mu - m \right) \psi(x) =0 \,,
\quad x^\mu = (t, \vec r) \,,
\end{equation}
where the matrices are used in the standard (Dirac) representation,
\begin{equation}
\gamma^0 =\beta = \left( \begin{array}{cc} \bm{1}_{2\times 2} & 0 \\
0 & -\bm{1}_{2\times 2} \\
\end{array} \right) \,,
\quad
\gamma^i = \left( \begin{array}{cc} 0 & \sigma^i \\ -\sigma^i & 0  \\
\end{array} \right) \,,
\quad
\gamma^5 = \left( \begin{array}{cc} 0 & \bm{1}_{2\times 2} \\
\bm{1}_{2\times 2} & 0  \\
\end{array} \right) \,.
\end{equation}
We define $\alpha^i = \gamma^0 \gamma^i$ and $\beta = \gamma^0$,
with Latin as opposed to Greek superscripts indicating the 
spatial components. The Lagrangian density is $\calL = \overline\psi(x) 
\left( \ii \gamma^\mu \partial_\mu - m \right) \psi(x)$,
and the Hamiltonian is Hermitian,
$H = \vec \alpha \cdot \vec p + \beta \, m$.
The fundamental positive-energy plain-wave eigenspinors are given by 
$\psi(x) = u_\pm(\vec k)  \, \ee^{-\ii k \cdot x}$ with 
$k \cdot x = E \,t - \vec k \cdot \vec r$ and 
\begin{equation}
\label{UU1}
u_+(\vec k) =
\left( \begin{array}{c}
\sqrt{\dfrac{E + m}{2 \, m}} \; a_+(\vec k) \\[2.33ex]
\sqrt{\dfrac{E - m}{2 \, m}} \; a_+(\vec k) 
\end{array} \right) \,,
\quad
u_-(\vec k) = \left( \begin{array}{c}
\sqrt{\dfrac{E + m}{2 \, m}} \; a_-(\vec k) \\[2.33ex]
-\sqrt{\dfrac{E - m}{2 \, m}} \; a_-(\vec k)
\end{array} \right) \,.
\end{equation}
We here use the covariant normalization 
$\overline u_\sigma(\vec k) \, u_\sigma(\vec k) = 
u^\plus_\sigma(\vec k) \, \gamma^0 \, u_\sigma(\vec k) = 1$,
which is different from 
Chap.~23 of Ref.~\cite{BeLiPi1982vol4} and
Sec.~2.8.1 on p.~28 of Ref.~\cite{GiKi2007}.
The energy is $E = \sqrt{ \vec k^2 + m^2}$.
Negative-energy eigenstates of the Dirac equation
in the helicity basis are given as
$\psi(x) = v_\pm(\vec k)  \, \ee^{\ii k \cdot x}$ with 
\begin{equation}
\label{VV1}
v_+(\vec k) = \left( \begin{array}{c}
-\sqrt{\dfrac{E - m}{2 \, m}} \; a_+(\vec k) \\[2.33ex]
-\sqrt{\dfrac{E + m}{2 \, m}} \; a_+(\vec k)
\end{array} \right) \,,
\quad
v_-(\vec k) = \left( \begin{array}{c}
-\sqrt{\dfrac{E - m}{2 \, m}} \; a_-(\vec k) \\[2.33ex]
\sqrt{\dfrac{E + m}{2 \, m}} \; a_-(\vec k)
\end{array} \right) \,.
\end{equation}
The normalization is $\overline v_\sigma(\vec k) \, v_\sigma(\vec k) = -1$.
The $a_\sigma(\vec k)$ are the fundamental 
helicity spinors, where the quantum number $\sigma = \pm$ is equal 
to the helicity for positive-energy states, and 
equal to the opposite helicity for negative-energy states.
In spherical coordinates, we have
\begin{equation}
\label{aaa}
a_+(\vec k) = \left( \begin{array}{c}
\cos\left(\frac{\theta}{2}\right) \\[0.33ex]
\sin\left(\frac{\theta}{2}\right) \, \ee^{\ii \, \varphi} \\
\end{array} \right) \,,
\quad
a_-(\vec k) = \left( \begin{array}{c}
-\sin\left(\frac{\theta}{2}\right) \, \ee^{-\ii \, \varphi} \\[0.33ex]
\cos\left(\frac{\theta}{2}\right) \\
\end{array} \right) \,.
\end{equation}
Here, $\theta$ and $\varphi$ are the polar and azimuthal angles
of the wave vector $\vec k$; they of course do 
{\em not} depend on the coordinate vector $\vec r$.
The eigenspinors fulfill the following projector sum rules,
\begin{equation}
\label{r1}
\sum_\sigma u_\sigma(\vec k) \otimes {\overline u}_\sigma(\vec k) =
\frac{\cancel{k} + m}{2 \, m} \,,
\quad
\sum_\sigma v_\sigma(\vec k) \otimes
{\overline v}_\sigma(\vec k) =
\frac{\cancel{k} - m}{2 \, m} \,.
\end{equation}

\begin{figure*}[t!]
\begin{center}
\includegraphics[width=0.57\linewidth]{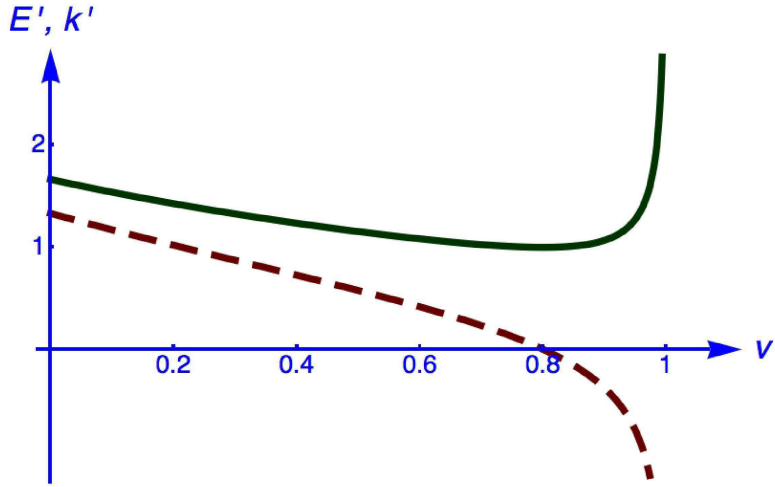} 
\caption{\label{fig1} Illustration of Eq.~\eqref{lorentz}.
The Lorentz-transformed energy $E'$ is plotted (solid curve) 
for $u = 0.8$, as seen in a frame of reference moving 
parallel to the $z$ axis at velocity $v$ (measured as a 
fraction of the speed of light). The energy attains a 
normalized value $E' = 1$ when the
particle is at rest in the moving frame (i.e., for $v = u$).
For $v > u$, the particle is overtaken and gains energy in the 
moving frame, while the $z$ component of the Lorentz-transformed
momentum $k'$ changes sign (dashed curve).}
\end{center}
\end{figure*}

\begin{figure*}[t!]
\begin{center}
\includegraphics[width=0.57\linewidth]{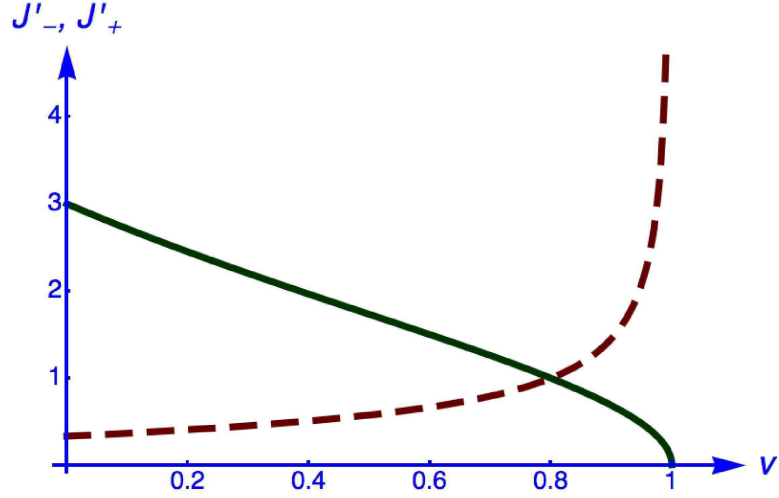}
\caption{\label{fig2} Illustration of Eq.~\eqref{flip2}.
The Lorentz-transformed current amplitude $J'_-$,
which corresponds to an (initially)
left-handed neutrino, is gradually sterilized
as the neutrino is overtaken (solid curve), and the helicity flips
(we use a value of $u=0.8$). Conversely, the current amplitude $J'_+$,
for an initially right-handed neutrino (dashed curve)
is gradually de-sterilized as the neutrino is overtaken.
The two curves cross at $v=u$, which is the moment
where the neutrino is being passed and the helicity flips.
Neither the left-handed nor the right-handed 
standard Dirac neutrino are ever completely sterile,
although $J'_{-} \to 0$ as $v \to 1$.
Furthermore, because the mass of the neutrino is 
Lorentz invariant, both helicity states correspond to 
very light fermions with the same mass as the left-handed neutrinos.}
\end{center}
\end{figure*}

We now consider in detail the transformation of the 
bispinor wave function under a Lorentz boost, which we 
denote by a change in the reference frame from the unprimed 
to the primed coordinate system. Namely,
under a Lorentz transformation $\Lambda$ with spinor 
representation $S(\Lambda)$, the bispinor wave function transforms as 
$\psi'(x') = S(\Lambda) \, \psi(x)$, with
\begin{align}
\psi(x) =& \; u_\sigma(\vec k) \, \exp(-\ii k \cdot x) 
\; \to \;
\psi'(x') = S(\Lambda) \, \psi(x)
= u_{\sigma'}(\vec k') \, \exp(-\ii k' \cdot x') \,,
\nonumber\\[0.133ex]
u_{\sigma'}(\vec k') =& \; S(\Lambda) \, u_\sigma(\vec k) \,.
\end{align}
The Dirac representation is recovered in the moving frame,
\begin{equation}
\gamma'^\mu = {\Lambda^\mu}_\nu  \, S(\Lambda) \,
\gamma^\nu \, S(\Lambda)^{-1} = \gamma^\mu \,,
\quad
{\Lambda^\mu}_\nu \, \gamma^\nu = 
S(\Lambda)^{-1} \, \gamma^\mu \, S(\Lambda) \,,
\end{equation}
where the latter relation is obtained from the former 
by the interchange $\mu \leftrightarrow \nu$ and 
multiplication by $S(\Lambda)^{-1}$ from the right and 
by $S(\Lambda)$ from the left.
We briefly verify that this is consistent with the
transformation of the theorem
$\bar u_\sigma(k) \, \gamma^\mu \, u_\sigma(k) = k^\mu/m$
under a Lorentz boost,
\begin{align}
\bar u_{\sigma'}(k') \, \gamma^\mu \, u_{\sigma'}(k') =& \;
\bar u_{\sigma}(k) \, S(\Lambda)^{-1} \, \gamma^\mu \, S(\Lambda) \, u_{\sigma}(k) 
\nonumber\\[0.133ex]
=& \; {\Lambda^\mu}_\nu \, \bar u_\sigma(k) \, \gamma^\nu \, u_\sigma(k) 
= {\Lambda^\mu}_\nu \, \frac{k^\nu}{m} = \frac{k'^{\mu}}{m} \,.
\end{align}
In regard to the elusive helicity flip~\cite{GoGo2011,Fe1998q76,Ho1998q76}, it is 
worthwhile to consider the current amplitude
\begin{equation}
\label{Jdef}
J_\sigma^\mu = \overline u_\sigma( \vec k ) \, \gamma^\mu \, (1-\gamma^5) \,
u_\sigma( \vec k ) \,,
\end{equation}
which describes the forward scattering of a positive-energy
neutrino in the helicity state $\sigma$ and with wave vector
$\vec k$. Under a Lorentz transformation ${\Lambda^\mu}_\nu$
with corresponding spinor transformation $S(\Lambda)$, this
amplitude transforms into
\begin{equation}
J'^\mu_\sigma = {\Lambda^\mu}_\nu \, \overline u_\sigma( \vec k ) \, 
\gamma^\nu (1-\gamma^5) \, u_\sigma(k)
=
\overline u_{\sigma'}( \vec k' ) \, \gamma^\mu \, (1-\gamma^5) \,
u_{\sigma'}( \vec k')
\end{equation}
For a Lorentz boost in the $z$ direction, we have
\begin{subequations}
\label{boost}
\begin{align}
{\Lambda^\mu}_\nu = & \;
\left( \begin{array}{cccc}
\cosh(\rho) & 0 & 0 & -\sinh(\rho) \\
0 & 1 & 0 & 0 \\
0 & 0 & 1 & 0 \\
-\sinh(\rho) & 0 & 0 & \cosh(\rho) \\
\end{array} 
\right) \,,
\\[0.133ex]
S(\Lambda) =& \; \left( \begin{array}{cccc}
\cosh(\tfrac12 \, \rho) & 0 & -\sinh(\tfrac12 \, \rho) & 0 \\
0 & \cosh(\tfrac12 \, \rho) & 0 & \sinh(\tfrac12 \, \rho) \\
-\sinh(\tfrac12 \, \rho) & 0 & \cosh(\tfrac12 \, \rho) & 0 \\
0 & \sinh(\tfrac12 \, \rho) & 0 & \cosh(\tfrac12 \, \rho) 
\end{array} 
\right) \,,
\end{align}
\end{subequations}
where $\rho$ is the rapidity and 
$\gamma = 1/\sqrt{1-v^2} = \cosh(\rho)$, 
whereas $\gamma \, v = \sinh(\rho)$.
We note that $S(\Lambda)$ acts in the space of 
bispinors, whereas ${\Lambda^\mu}_\nu $ acts in the space
of Lorentz vectors, and that the matrix representation of 
$S(\Lambda)$ is tied to the Dirac representation which we use for the 
$\gamma$ matrices. Surprisingly, 
explicit expressions for spinor Lorentz boosts in the 
Dirac representation of the Dirac algebra seem to be
rather scarce in the literature; our formula is in agreement 
with the discussion in Ref.~\cite{weyllorentz} and 
with Eq.~(2.74) in Sec.~2.4.1 
on p.~15 of Ref.~\cite{GiKi2007}. Note that $S(\Lambda)$
would read differently in the chiral
representation of the Dirac matrices. Irrespective
of the matrix representation of the Dirac algebra, we have 
\begin{equation}
\label{boostgen}
S(\Lambda) = \exp\left( - \tfrac{\ii}{4} \, 
\sigma^{\alpha \, \beta} \, 
\omega_{\alpha \, \beta} \right) \,,
\quad
\sigma^{\alpha \, \beta}  = 
\tfrac{\ii}{2} \, [\gamma^\alpha, \gamma^\beta] \,,
\quad
\omega_{03} = -\omega_{30} = -\rho \,,
\end{equation}
while all other elements $\omega_{\mu\nu}$ vanish.
We assume that $\vec k = k \, \hat e_z$ points into the 
positive $z$ direction,
while the Lorentz-boosted $\vec k' = k' \, \hat e_z$
may either point into the positive or negative $z$ direction,
depending on whether $k'$ is positive or negative.
Energy $E'$ and $z$ component of the momentum $k'$ are given as follows,
\begin{equation}
\label{lorentz}
E = \frac{m}{\sqrt{1 - u^2}} \,,
\;
k = \frac{m \, u}{\sqrt{1 - u^2}} \,,
\;
E' = \frac{m (1 - u \,v )}{\sqrt{1 - v^2} \, \sqrt{1-u^2}} \,,
\;
k' = \frac{m (u - v )}{\sqrt{1 - v^2} \, \sqrt{1-u^2}} \,.
\end{equation}
The Lorentz-boosted $k'$ is negative for $v > u$, as it should.
Let us now investigate the helicity
$\sigma'$ of the Lorentz-transformed spinor.
We observe that, according to Eq.~\eqref{aaa},
setting $\theta = 0$ and $\theta = \pi$,
\begin{subequations}
\label{aaa2}
\begin{align}
a_+(|\vec k| \, \hat e_z) =& \; \left( \begin{array}{c}
1 \\[0.33ex] 0 \\ \end{array} \right) \,,
\quad
a_-(|\vec k| \, \hat e_z) = \left( \begin{array}{c}
0 \\[0.33ex] 1 \\ \end{array} \right) \,,
\\[0.133ex]
a_+(-|\vec k| \, \hat e_z) =& \; \left( \begin{array}{c}
0 \\[0.33ex] 1 \\ \end{array} \right) \,,
\quad
a_-(-|\vec k| \, \hat e_z) = \left( \begin{array}{c}
1 \\[0.33ex] 0 \\ \end{array} \right) \,.
\end{align}
\end{subequations}
Here, we have chosen the constant, unobservable phase 
$\exp(\pm \ii \varphi)$ of the helicity spinor equal 
so that the nonvanishing entry is equal to unity in 
each case. The helicity flip in the spinor wave function thus finds
a natural mathematical correspondence: Once the 
$z$ component of the vector $\vec k$ flips, the interpretation of the 
fundamental helicity spinor also flips. 
Acting with the spinor transformation $S(\Lambda)$ onto the 
bispinor, it is then possible to verify, after some algebraic
manipulation,
that the bispinors given in Eq.~\eqref{UU1} transform as follows,
\begin{subequations}
\label{flip1}
\begin{align}
u_+(\vec k) \to & \; S(\Lambda) \, u_+(\vec k) = u_+(\vec k') = 
\left( \begin{array}{c}
\sqrt{\dfrac{E' + m}{2 \, m}} \; a_+(\vec k') \\[2.33ex]
\sqrt{\dfrac{E' - m}{2 \, m}} \; a_+(\vec k') 
\end{array} \right) \,, 
\quad v < u \,,
\\[0.133ex]
u_+(\vec k) \to & \; S(\Lambda) \, u_+(\vec k) = u_-(\vec k') = 
\left( \begin{array}{c}
\sqrt{\dfrac{E' + m}{2 \, m}} \; a_-(\vec k') \\[2.33ex]
-\sqrt{\dfrac{E' - m}{2 \, m}} \; a_-(\vec k') 
\end{array} \right) \,, 
\quad
v > u \,,
\\[0.133ex]
u_-(\vec k) \to & \; S(\Lambda) \, u_-(\vec k) = u_-(\vec k') = 
\left( \begin{array}{c}
\sqrt{\dfrac{E' + m}{2 \, m}} \; a_-(\vec k') \\[2.33ex]
-\sqrt{\dfrac{E' - m}{2 \, m}} \; a_-(\vec k')
\end{array} \right) \,,
\quad v < u \,,
\\[0.133ex]
u_-(\vec k) \to & \; S(\Lambda) \, u_-(\vec k) = u_+(\vec k') = 
\left( \begin{array}{c}
\sqrt{\dfrac{E' + m}{2 \, m}} \; a_+(\vec k') \\[2.33ex]
\sqrt{\dfrac{E' - m}{2 \, m}} \; a_+(\vec k')
\end{array} \right) \,,
\quad v > u \,.
\end{align}
\end{subequations}
These equations explicitly verify the presence of the helicity flip,
within the bispinor representation of the Lorentz group.
Our analysis is in agreement with the brief discussion
presented in the text following Eq.~(2.331) on p.~46 of Ref.~\cite{GiKi2007}.
Helicity flips upon a transformation into a rotating frame of reference
are discussed in~\ref{appa}. For the two helicities $\sigma = \pm$,
and the time-like component $\mu=0$, 
defining ${J'}_\sigma \equiv {J'}_\sigma^{\mu=0}$, we obtain 
from the Lorentz boost,
\begin{subequations}
\label{flip2}
\begin{align}
{J'}_- =& \; \left\{ \begin{array}{cc} 
\dfrac{E'}{m} \, 
\left( 1 + \sqrt{1 - \dfrac{m^2}{E'^2}} \right) \,, & \quad v < u \,, \\[2.133ex]
\dfrac{E'}{m} \, 
\left( 1 - \sqrt{1 - \dfrac{m^2}{E'^2}} \right) \,, & \quad v > u \,,
\end{array} \right. 
\\[0.133ex]
{J'}_+ =& \; \left\{ \begin{array}{cc} 
\dfrac{E'}{m} \, 
\left( 1 - \sqrt{1 - \dfrac{m^2}{E'^2}} \right) \,, & \quad v < u \,, \\[2.133ex]
\dfrac{E'}{m} \, 
\left( 1 + \sqrt{1 - \dfrac{m^2}{E'^2}} \right) \,, & \quad v > u \,.
\end{array} \right. 
\end{align}
\end{subequations}
A discussion is in order. The functional form of the 
Lorentz-transformed amplitudes ${J'}_-$ and ${J'}_+$ changes at
$v = u$, which in turn corresponds to the velocity necessary to 
overtake the neutrino. The $z$ component of the momentum 
changes sign at this velocity, and this is manifest in Eq.~\eqref{flip1},
where the change in the helicity quantum number is manifest.
We recall that $\vec k' = k' \, \hat e_z$, where $k'$ is given in 
Eq.~\eqref{lorentz}; it changes direction at $v = u$. Indeed,
the neutrino is at rest in the moving frame for $v = u$.

The asymptotics of ${J'}_-$ for $v < u$, for large energy $E'$, go as
${J'}_- \to 2 E'/m$, in agreement with the left-handed (negative) helicity 
of the ``active'' neutrino, while the helicity flip at $v=u$ implies
that  ${J'}_- \to m/(2 \, E')$ for $v > u$, for large energy $E'$.
In the moving frame, the overtaken neutrino thus becomes gradually 
sterile, and the change in the functional form in Eq.~\eqref{flip2}
is consistent with a smooth behaviour of the current amplitude
near the point where the neutrino actually is being passed ($v = u$).
The suppression factor upon comparing the two asymptotic 
forms thus reads as $m^2/(4 \, E'^2)$, where $E'$ is given 
according to Eq.~\eqref{lorentz} (see also Fig.~\ref{fig1}).

Conversely, the asymptotics of ${J'}_+$ for $v < u$, for large energy $E'$,
go as ${J'}_+ \to m/(2 \, E')$ and are thus suppressed, in agreement with the
right-handed (negative) helicity of the ``active'' neutrino, while the helicity
flip at $v=u$ implies that  ${J'}_+ \to 2 E'/m$ for $v > u$, for large energy
$E'$.  In the moving frame, the overtaken neutrino is ``de-sterilized''.
Despite this parametric suppression at high energies, the right-handed Dirac
neutrinos are thus not completely sterile.  This consideration quantifies and
illustrates the ``helicity reversal question'' which underlies the
considerations of Refs.~\cite{Fe1998q76,Ho1998q76,GoGo2011}, and
Fig.~\ref{fig2} clearly demonstrates how the helicity reversal and the
concomitant gradual suppression of the weak interaction occurs as the velocity
of the reference frame $v$ approaches the speed $u$ of the neutrino.

%
%
\section{Majorana Particles and Helicity Flip}
\label{sec3}

Recently, Majorana-particle-like excitations have been observed in solid-state
physics~\cite{MoEtAl2012}.  Also, a quantum propagation
algorithm~\cite{CaEtAl2011} for Majorana particle wave packets has been
devised, which takes the particle-antiparticle symmetry of the Majorana wave
function into account.  These questions are nontrivial because the Majorana
particle formally follows the Dirac equation, but with the additional
constraint that the wave function is charge-conjugation invariant, i.e.~it
fulfills the condition $\psi^\calC = C \, {\overline \psi}^{\rm T}$ with $C =
\ii \, \gamma^2 \, \gamma^0$ (in the Dirac representation which we use here).

At variance with Ref.~\cite{KaPeGB1989},
we here consider a space-time metric has the signature
$(+,-,-,-)$ so that $k \cdot x = E \, t - \vec k \cdot \vec r$.
The Majorana fermion is described by the 
field operator [see Eq.~(3.25) on p.~47 of Ref.~\cite{KaPeGB1989}],
\begin{equation}
\label{fk}
{\hat \psi}(x) = \sum_{\vec k, \sigma} 
\sqrt{\frac{m}{E \, V}} \,
\left[ 
f_\sigma(\vec k) \, u_\sigma(\vec k) \, \ee^{- \ii k \cdot x} +
\lambda \, 
f^+_{-\sigma}(\vec k) \, v_{-\sigma}(\vec k) \, \ee^{\ii k \cdot x} 
\right]
\end{equation}
where the operator $f_\sigma(\vec k)$ annihilates a 
Majorana fermion, whereas $f^+_\sigma(\vec k) $ creates one.
The peculiar $-\sigma$ subscript in the antiparticle 
contribution to ${\hat \psi}(x) $ stems from the 
fact that in our conventions, $\sigma$ is a quantum 
number which is equal to the helicity for particles,
whereas it is equal to the negative of the helicity 
for antiparticles [see the text following Eq.~\eqref{VV1}
and the discussion in~\ref{appc}, especially Eq.~\eqref{CCinv}].
The creation phase factor $\lambda = \exp(\ii \, \varphi)$ has unit 
modulus~\cite{KaPeGB1989}.
Up to a phase factor, the Majorana field operator is equal 
to its own charge conjugate $\calC \, {\hat \psi}(x) \, \calC^{-1}$ 
[see Eq.~(3.29) on p.~51 of Ref.~\cite{KaPeGB1989}].
This form of the field operator ensures that the 
time-ordered vacuum expectation value reproduces the 
Feynman form $\ii/(\cancel{k} - m + \ii \, \epsilon)$ 
in momentum space (see p.~67 of Ref.~\cite{KaPeGB1989}).

\begin{figure*}[t!]
\begin{center}
\includegraphics[width=0.57\linewidth]{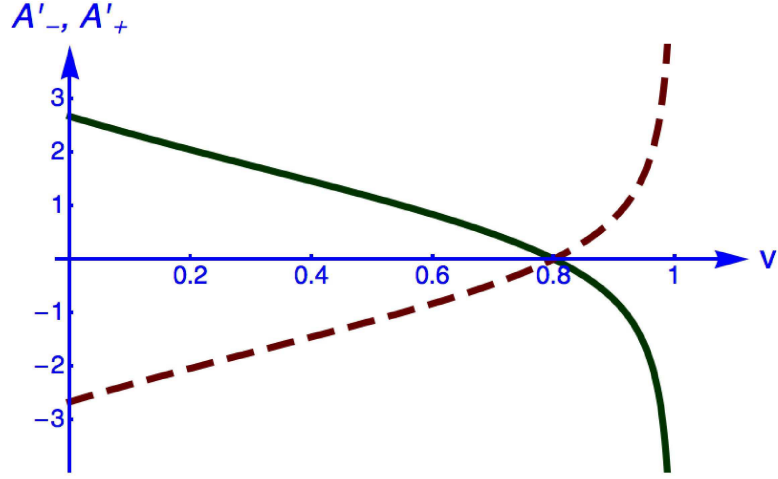}
\caption{\label{fig3} The axial-vector matrix element
$A'_\sigma$ for Majorana neutrinos 
replaces the current $J'_\sigma$ plotted
for Dirac particles in Fig.~\ref{fig2}
(the solid curve is for $A'_{-}$, the dashed one for $A'_{+}$).
According to Eq.~\eqref{flip3}, 
the helicity of the Majorana neutrino flips
(we again use a value of $u=0.8$ for the velocity,
the neutrino is overtaken for $v > u$).
However, there is no sterilization as one boosts
past the Majorana neutrino.}
\end{center}
\end{figure*}

Because a Majorana fermion is equal to its own antiparticle, the 
vector and axial vector currents corresponding to 
single-particle state of the Majorana inevitably acquire
contributions from the anti-particle solutions.
The correspondence is as follows.
We consider the matrix elements
$\left< \nu_f \left| \hat{\overline \psi} \; 
\gamma^\mu \; \hat{\psi} \; \right| \nu_i \right>$
and $\left< \nu_f \left| \hat{\overline \psi} \; 
\gamma^\mu \, \gamma^5 \; \hat{\psi} \; \right| \nu_i \right>$
where $\hat{\psi} $ is the Majorana field operator and 
$| \nu_i \rangle = f^+_\sigma(\vec k) \, | 0 \rangle$
is a one-particle state lifted from the vacuum
(the subscript $i$ and $f$ denote initial and final states).
According to the derivation presented on pp.~61--62 
of Ref.~\cite{KaPeGB1989}, one can show that 
\begin{align}
\label{generated}
\left< \nu_f \left| \hat{\overline \psi} \; 
\gamma^\mu \; \hat{\psi} \; \right| \nu_i \right> \propto & \;
{\overline u}_f \; \gamma^\mu \; u_i - 
{\overline v}_i \; \gamma^\mu \; v_f \equiv V^\mu \,,
\\[0.133ex]
\left< \nu_f \left| \hat{\overline \psi} \; 
\gamma^\mu \, \gamma^5 \; \hat{\psi} \; \right| \nu_i \right> \propto & \;
- \left( {\overline u}_f \; \gamma^\mu \, \gamma^5 \; u_i - 
{\overline v}_i \; \gamma^\mu \, \gamma^5 \; v_f \right) \equiv A^\mu \,,
\end{align}
We here define the vector current and axial vector 
current matrix elements $V^\mu$ and $A^\mu$, respectively.
The negative prefactor in the definition of $A^\mu$ 
will find an immediate explanation.
For forward scattering, the matrix elements
corresponding to $J^\mu_\sigma$ defined in 
Eq.~\eqref{Jdef} are thus as follows,
\begin{subequations}
\begin{align}
\label{VAdef}
V_\sigma^\mu = & \;
\overline u_\sigma( \vec k ) \; \gamma^\mu \; u_\sigma( \vec k ) -
{\overline v}_{-\sigma}(\vec k ) \; \gamma^\mu \; v_{-\sigma}( \vec k ) \,,
\\[0.133ex]
A_\sigma^\mu = & \;
- \left( 
\overline u_\sigma( \vec k ) \; \gamma^\mu \, \gamma^5 \; u_\sigma( \vec k ) -
{\overline v}_{-\sigma}(\vec k ) \; \gamma^\mu \, \gamma^5 \; v_{-\sigma}( \vec k ) 
\right) \,.
\end{align}
\end{subequations}
The charge conjugation matrix $C = \ii \, \gamma^2 \, \gamma^0$ 
has the properties,
\begin{subequations}
\begin{align}
C =& \; \ii \, \gamma^2 \, \gamma^0 \,,
\qquad C^2 = -{\bf 1}_{4 \times 4} \,, \qquad C^{-1} = -C \,,
\\[0.133ex]
C^{-1} \; \gamma^\mu \; C =& \; -\left( \gamma^\mu \right)^{\rm T} \,,
\qquad
C^{-1} \; \gamma^5 \; C = \left( \gamma^5 \right)^{\rm T} \,,
\\[0.133ex]
v_{-\sigma}(\vec k) =& \; C \, {\overline u}_\sigma^{\rm T}(\vec k) \,,
\qquad
{\overline v}_{-\sigma}(\vec k) = - u^{\rm T}_\sigma(\vec k) \, C^{-1} \,,
\end{align}
\end{subequations}
where ${\bf 1}_{4 \times 4}$ is the four-dimensional unit matrix.
Hence, for the vector current, we have
\begin{align}
{\overline v}_{-\sigma}(\vec k) \; \gamma^\mu \; v_{-\sigma}(\vec k)
=& \; - u_\sigma^{\rm T}(\vec k) \; C^{-1} \; \gamma^\mu \; C \; 
{\overline u}_\sigma^{\rm T}(\vec k)
= {\overline u}_\sigma(\vec k) \; \gamma^\mu \; u_\sigma(\vec k) \,,
\end{align}
whereas for the axial vector current,
\begin{align}
{\overline v}_{-\sigma}(\vec k) \; \gamma^\mu \, \gamma^5 \; v_{-\sigma}(\vec k)
=& \; - u_\sigma^{\rm T}(\vec k) \; C^{-1} \; \gamma^\mu \; C \;
C^{-1} \; \gamma^5 \; C \; {\overline u}_\sigma(\vec k) 
\nonumber\\[2ex]
=& \; - u_\sigma^{\rm T}(\vec k) \; \left( -\gamma^\mu \right)^{\rm T} \;
\left( \gamma^5 \right)^{\rm T} \; {\overline u}_\sigma(\vec k) 
= u_\sigma^{\rm T}(\vec k) \; \left( \gamma^5 \; \gamma^\mu \right)^{\rm T} \;
{\overline u}_\sigma^{\rm T}(\vec k)
\nonumber\\[2ex]
=& \; - u_\sigma^{\rm T}(\vec k) \; \left( \gamma^\mu \; \gamma^5 \right)^{\rm T} \;
{\overline u}_\sigma^{\rm T}(\vec k)
= - {\overline u}_\sigma(\vec k) \; \gamma^\mu \; \gamma^5 \; u_\sigma(\vec k) \,.
\end{align}
So, we have
\begin{align}
\label{VAres}
V_\sigma^\mu = 0 \,,
\qquad
A_\sigma^\mu = 
-2 \, \overline u_\sigma( \vec k ) \; \gamma^\mu \, \gamma^5 \; u_\sigma( \vec k )  \,,
\end{align}
a result which confirms the common wisdom that 
Majorana particles have no vector current (see p.~60 
of Ref.~\cite{KaPeGB1989}).
The $V$-$A$ coupling for Majorana particles is 
exclusively carried by the axial component.
The prefactor in $A_\sigma^\mu$ is chosen so that, 
in the correspondence of Eqs.~\eqref{Jdef} and~\eqref{VAres},
\begin{equation}
\label{corresp}
J_\sigma^\mu \;\;\; \mbox{(for Dirac neutrinos)} 
\;\; \Leftrightarrow \;\;
A_\sigma^\mu \;\;\; \mbox{(for Majorana neutrinos)} \,.
\end{equation}
The amplitudes $J_\sigma^\mu  \Leftrightarrow A_\sigma^\mu$ 
are matched in the high-energy limit 
where the helicity eigenstates approximate the chirality 
eigenstates and are ``almost'' eigenstates of $\gamma^5$.
In analogy to Eq.~\eqref{flip2}, we define
$A'_\sigma = A'^{\mu = 0}_\sigma$ for the time-like component. The following 
results are easily obtained for the 
Lorentz-transformed axial Majorana currents,
\begin{subequations}
\label{flip3}
\begin{align}
{A'}_- =& \; \left\{ \begin{array}{cc} 
\dfrac{ 2 E'}{m} \, \sqrt{1 - \dfrac{m^2}{E'^2}} \,, & \quad v < u \,, \\[2.133ex]
- \dfrac{ 2 E'}{m} \, \sqrt{1 - \dfrac{m^2}{E'^2}} \,, & \quad v > u \,,
\end{array} \right. 
\\[0.133ex]
{A'}_+ =& \; \left\{ \begin{array}{cc} 
- \dfrac{ 2 E'}{m} \, \sqrt{1 - \dfrac{m^2}{E'^2}} \,, & \quad v > u \,, \\[2.133ex]
\dfrac{ 2 E'}{m} \, \sqrt{1 - \dfrac{m^2}{E'^2}} \,, & \quad v > u \,.
\end{array} \right.  
\end{align}
\end{subequations}
These results confirm the Dirac--Majorana confusion
theorem~\cite{KaSh1981,LiWi1982,Ka1982}: For $E' \gg m$, we have
$J'_\sigma \approx A'_\sigma \approx 2 E'/m$, but this holds only for
Lorentz transformations into frames which do not ``overtake'' the neutrino ($v
< u$).  As we ``overtake'' the Majorana neutrino, the left-handed Majorana
neutrino transforms into a right-handed Majorana neutrino, which however,
because the Majorana is its own anti-particle, needs to interact as if it were
a right-handed Majorana anti-neutrino, hence it is not sterile.  This is
illustrated in Fig.~\ref{fig3}.  Majorana neutrinos are not sterilized by
overtaking them in a gedanken experiment on a highway.  To put the conclusion
into perspective, a left-handed neutrino incident on a target at rest in the
frame in which it is born will generate leptons, but when one boosts past it and
the helicity flips, if it is a Majorana particle,
then an interaction with a target at rest in
the boosted frame will generate antileptons.  This is consistent with the
lepton-number nonconservation induced into the Standard Model by Majorana
neutrinos (an illustrative discussion is found in Sec.~6.1.1 on pp.~184--185 of
Ref.~\cite{GiKi2007}).

%
%
\section{Generalized Dirac Equation and Helicity Suppression}
\label{sec4}

In order to start the discussion of generalized 
Dirac equations with exotic dispersion relations,
we first approach the light cone by considering 
massless Dirac fermions, which are always in a helicity eigenstate,
if they are in an energy eigenstate.
The massless Dirac equation and its Hamiltonian $H_0$ read, quite simply, 
$\ii \gamma^\mu \,\partial_\mu \; \psi(x) = 0$,
and $H_0 = \vec\alpha \cdot \vec p$, respectively,
where we again use the standard Dirac representation, and 
the $\vec\sigma$ are the $2 \times 2$ Pauli matrices.
The massless Dirac Hamiltonian is a $4 \times 4$ matrix
and it is simultaneously Hermitian as well as 
pseudo-Hermitian or ``$\gamma^5$--Hermitian''~\cite{Pa1943,BeBo1998,%
BeBoMe1999,BeWe2001,
Mo2002i,Mo2002ii,Mo2003npb,JeSuZJ2009prl,JeSuZJ2010,GaLa2009},
\begin{equation}
H_0 = H_0^\plus \,,
\quad
H_0 = \gamma^5 \; H_0^\plus \; \gamma^5 \,.
\end{equation}
One chooses a plane-wave ansatz of the form
$\psi(x) = u_{\sigma}(\vec k) \, \exp(-\ii \, k \cdot x)$
for particles and 
$\psi(x) = v_{\sigma}(\vec k) \, \exp(\ii \, k \cdot x)$
for antiparticles, where $k \cdot x = |\vec k| \, t - \vec k \cdot \vec r$.
According to Sec.~2.4.3 of Ref,~\cite{ItZu1980},
the canonical choice for the phases of
the fundamental eigenspinors is as follows,
\begin{subequations}
\label{C}
\begin{align}
u_+(\vec k) = & \;
\frac{1}{\sqrt{2}}
\left( \begin{array}{c}
a_+(\vec k) \\[0.33ex]
a_+(\vec k) \\
\end{array} \right) \,,
\quad
u_-(\vec k) =
\frac{1}{\sqrt{2}}
\left( \begin{array}{c}
a_-(\vec k) \\[0.33ex]
-a_-(\vec k) \\
\end{array} \right) \,,
\\[0.133ex]
v_+(\vec k) = & \;
\frac{1}{\sqrt{2}}
\left( \begin{array}{c}
-a_+(\vec k) \\[0.33ex]
-a_+(\vec k) \\
\end{array} \right) \,,
\quad
v_-(\vec k) =
\frac{1}{\sqrt{2}}
\left( \begin{array}{c}
-a_-(\vec k) \\[0.33ex]
a_-(\vec k) \\
\end{array} \right) \,.
\end{align}
\end{subequations}
The zero-mass solutions simultaneously fulfill the following 
projector identities,
\begin{subequations}
\label{massless_sum}
\begin{align}
\label{massless_sum1}
\sum_\sigma  \; 2 \, |\vec k| \, u_\sigma(\vec k) \otimes
\overline u_\sigma(\vec k) = & \;
\sum_\sigma \; 2 \, |\vec k| \, v_\sigma(\vec k) \otimes
\overline v_\sigma(\vec k) =
\cancel{k} \,,
\\[2ex]
\label{massless_sum2}
\sum_\sigma  \; 2 \, |\vec k| \, (-\sigma) \, u_\sigma(\vec k) \otimes
\overline u_\sigma(\vec k) \, \gamma^5 = & \;
\sum_\sigma \; 2 \, |\vec k| \, (-\sigma) \, v_\sigma(\vec k) \otimes
\overline v_\sigma(\vec k) \, \gamma^5 =
\cancel{k} \,,
\end{align}
\end{subequations}
as one can check by an explicit calculation.
We note, in particular, that it is 
impossible to ``choose a covariant normalization''
in the massless case; one always has
$\overline u_\sigma(\vec k) \, u_\sigma(\vec k) = 
u^\plus_\sigma(\vec k) \, \gamma^0 \, u_\sigma(\vec k) = 0$
and $\overline v_\sigma(\vec k) \, v_\sigma(\vec k) = 0$
for the massless spinors given in Eq.~\eqref{C},
hence, the prefactors $|\vec k|$ in Eq.~\eqref{massless_sum}.

The tachyonic generalized Dirac equation reads as 
\begin{equation}
\label{eqtachyon}
\left( \ii \gamma^\mu \partial_\mu - \gamma^5 m \right)
\psi(x) =0 \,,
\end{equation}
with a Lagrangian density 
$\calL = \overline\psi(x) \, \gamma^5 \,
\left( \ii \gamma^\mu \partial_\mu - \ii m \right)
\psi(x)$.
The corresponding Hamiltonian~\cite{ChHaKo1985,ChKoPoGa1992,Ko1993,ChKo1994,%
JeWu2012epjc,JeWu2012jpa,JeWu2013isrn}
is pseudo-Hermitian or ``$\gamma^5$--Hermitian'',
\begin{equation}
H = \vec \alpha \cdot \vec p + \beta \, \gamma^5 \, m \,,
\quad
H = \gamma^5 \, H^\plus \, \gamma^5 \,.
\end{equation}
The concept of pseudo-Hermiticity has been established
as a viable generalization of Hermiticity for 
quantum systems~\cite{Pa1943,BeBo1998,%
BeBoMe1999,BeWe2001,
Mo2002i,Mo2002ii,Mo2003npb,JeSuZJ2009prl,JeSuZJ2010},
with pseudo-Hermitian Hamiltonians describing systems
where the absorptive (``gain'') and 
dissipative (``loss'') terms are in 
equilibrium, and the resulting eigenenergy is real
rather than complex~\cite{NoLuJe2013epjp}.
Such systems are physically realized in so-called 
$\mathcal{P}\mathcal{T}$-symmetric waveguides~\cite{Jo2010optics}
and optical lattices~\cite{SzReBHSe2011}.
The application to a superluminal spin-$1/2$ particle
constitutes a candidate theory for a more fundamental 
application of the concept of pseudo-Hermiticity,
whose phenomenological relevance remains to be
tested experimentally.

The fundamental positive-energy and negative-energy 
eigenstates of the tachyonic Dirac equation
in the helicity basis are given as
$\psi(x) = u_\sigma(\vec k)  \, \ee^{\ii k \cdot x}$ and
$\psi(x) = v_\sigma(\vec k)  \, \ee^{\ii k \cdot x}$ with
with $k \cdot x = E \, t - \vec k \cdot \vec r$
and $E = \sqrt{ \vec k^2 - m^2}$.
The fundamental eigenspinors are obtained 
from Eqs.~\eqref{UU1} and~\eqref{VV1} by the simple
(formal) substitution $E \to |\vec k|$~(see Ref.~\cite{JeWu2013isrn}),
\begin{equation}
\label{UT1}
u_+(\vec k) =
\left( \begin{array}{c}
\sqrt{\dfrac{|\vec k| + m}{2 \, m}} \; a_+(\vec k) \\[2.33ex]
\sqrt{\dfrac{|\vec k| - m}{2 \, m}} \; a_+(\vec k) 
\end{array} \right) \,,
\quad
u_-(\vec k) = \left( \begin{array}{c}
\sqrt{\dfrac{|\vec k| + m}{2 \, m}} \; a_-(\vec k) \\[2.33ex]
-\sqrt{\dfrac{|\vec k| - m}{2 \, m}} \; a_-(\vec k)
\end{array} \right) \,,
\end{equation}
with the normalization
$\overline u_\sigma(\vec k) \, u_\sigma(\vec k) = \sigma$, and 
\begin{equation}
\label{VT1}
v_+(\vec k) = \left( \begin{array}{c}
-\sqrt{\dfrac{|\vec k| - m}{2 \, m}} \; a_+(\vec k) \\[2.33ex]
-\sqrt{\dfrac{|\vec k| + m}{2 \, m}} \; a_+(\vec k)
\end{array} \right) \,,
\quad
v_-(\vec k) = \left( \begin{array}{c}
-\sqrt{\dfrac{|\vec k| - m}{2 \, m}} \; a_-(\vec k) \\[2.33ex]
\sqrt{\dfrac{|\vec k| + m}{2 \, m}} \; a_-(\vec k)
\end{array} \right) \,,
\end{equation}
with the normalization
$\overline v_\sigma(\vec k) \, v_\sigma(\vec k) = -\sigma$.
The projector sums reads as follows,
\begin{equation}
\label{r2}
\sum_\sigma (-\sigma) \; u_\sigma(\vec k) \otimes
\overline u_\sigma(\vec k) \,\gamma^5 =
\frac{\cancel{k} - \gamma^5 \, m}{2 m} \,, 
\quad
\sum_\sigma (-\sigma) \; v_\sigma(\vec k) \otimes
\overline v_\sigma(\vec k) \,\gamma^5 =
\frac{\cancel{k} + \gamma^5 \, m}{2 m} \,. 
\end{equation}
We recall that $\sigma$ is a ``good'' quantum number and 
characterizes the helicity, not chirality.
Indeed, $\sigma$ is equal to the helicity for 
positive-energy states, and equal to minus the 
helicity for negative-energy states.
In the massless limit~\cite{JeWu2013isrn},
the ``second'' projector sum~\eqref{massless_sum2} is
recovered.

One might wonder about 
the ``mysterious'' factor $(-\sigma)$ in the sum rules.
If we postulate that the time-ordered product of 
field operators should give a propagator whose Fourier
transformation is the inverse of the Hamiltonian,
then we have to sum over the eigenspinors, as is done in 
any derivation of a Green function, and obtain the 
positive-energy and negative-energy projectors over the 
eigenspinors as a result of the spin sums~\cite{JeWu2013isrn}. 
So, we have to postulate that an expression of the 
form 
\begin{equation}
\label{r3}
\sum_{\rm spin} \mbox{(prefactor)} \times
\mbox{(tensor product of states)} \times 
\mbox{(scalar or pseudo-scalar matrix)} 
\end{equation}
must be equal to a positive-energy or negative-energy 
projector. Here, the prefactor can only come from the 
fundamental anticommutator of the field operator, which 
in turn can only involve the ``good'' quantum numbers. 
We postulate a relation of the form
\begin{equation}
\label{anticom}
\left\{ b_\sigma(k) , b^\plus_{\rho}(k') \right\} = 
f(\sigma, \vec k) \;
(2 \pi)^3 \, \frac{E}{m} \delta^3(\vec k - \vec k') \,
\delta_{\sigma\rho}\,,
\end{equation}
where the $b$ and $b^+$ annihilate and create 
particles of the respective quantum numbers,
and corresponding relations for the creators and annihilators 
of antiparticles.
The ``prefactor'' in Eq.~\eqref{r3} has to be
a function of ``good'' quantum 
numbers~\cite{JeWu2012epjc,JeWu2012jpa,JeWu2013isrn}
and therefore of the functional form $f(\sigma, \vec k)$,
\begin{equation}
\mbox{(prefactor)} = f(\sigma, \vec k) \,,
\end{equation}
where $f(\sigma, \vec k)$ enters Eq.~\eqref{anticom}.
For the further derivation of the tachyonic propagator, 
one simply generalizes the derivation that
leads from Eq.~(3.169) to Eq.~(3.170)
in the standard textbook~\cite{ItZu1980},
as it has been described in Ref.~\cite{JeWu2013isrn}.
In turn, an inspection of Eqs.~\eqref{r1} and~\eqref{r2}
shows that for the choices
\begin{equation}
f(\sigma, \vec k) = 1 \quad \mbox{(tardyonic choice)} \,,
\quad
f(\sigma, \vec k) = -\sigma \quad \mbox{(tachyonic choice)} \,,
\end{equation}
the sum rules~\eqref{r1} and~\eqref{r2} are fulfilled,
and the propagator may be calculated~\cite{JeWu2013isrn}.
If we accept the anticommutator relation~\eqref{anticom},
then the right-handed helicity state acquires
negative norm. This is seen as follows,
\begin{equation}
\label{negnorm}
\left< 1_{k, \sigma} | 1_{k, \sigma} \right> =
\left< 0 \left| b_\sigma(k) \,
b^+_\sigma(k) \right| 0 \right>
= \left< 0 \left| \left\{ b_\sigma(k),
b^+_\sigma(k) \right\} \right| 0 \right>
= (-\sigma) \, V \, \frac{E}{m} \,,
\end{equation}
where $V = (2 \pi)^3 \, \delta^3(\vec 0)$ is the normalization
volume in coordinate space. The norm
$\left< 1_{k, \sigma} | 1_{k, \sigma} \right>$ is negative for $\sigma = 1$.
Therefore, right-handed particle states
are excluded from the spectrum by a 
Gupta--Bleuler condition~\cite{JeWu2012epjc,JeWu2013isrn}.
This is analogous to virtual photons which can be 
scalar or longitudinal, but physical photons must be 
transverse (see also~\ref{appb}).
The helicity reversal question discussed in Sec.~\ref{sec2}
does not need to be addressed for (even infinitesimally) 
superluminal neutrinos because it is impossible to 
overtake them, starting from rest, by the laws of special relativity.

%
%
\section{Conclusions}
\label{sec5}

We have presented an overview of helicity suppression mechanisms for the
``wrong'' neutrino helicity states, under the assumption that the neutrino mass
eigenstates are either described by the standard Dirac equation (see
Sec.~\ref{sec2}), or by the tachyonic Dirac equation (see Sec.~\ref{sec4}).
The case of a Majorana neutrino is treated in Sec.~\ref{sec3}. For a Dirac
neutrino, incomplete sterilization occurs if one overtakes a left-handed
neutrino while speeding on a highway without speed limits. The same mechanism
predicts gradual de-sterilization upon overtaking right-handed Dirac neutrinos.
This phenomenon is graphically represented in Fig.~\ref{fig2}.  The weak
interactions of the ``wrong'' helicity states, for a standard Dirac neutrino,
are suppressed by a power of $m/E$, where $m$ is the neutrino mass and $E$ is
the energy scale.  In addition, one should point out that the (almost) sterile
``left-handed turned right-handed'' (after the passing maneuver) Dirac
neutrinos still have the same mass as before the overtaking: They represent
not-quite-sterile, light, right-handed states of the fermion field. Perhaps
even more counter-intuitive (but consistently described by a spinor Lorentz
transformation) is the gradual de-sterilization of a bunch of right-handed
neutrinos as one passes them on a highway (see also Fig.~\ref{fig2}).  

By contrast, for Majorana neutrinos, no sterilization occurs (see
Sec.~\ref{sec3}).  The antiparticle-component of the axial vector current,
which is always present for a Majorana particle, leads to a ``built-in''
transformation of the Majorana neutrino into its own antiparticle upon
overtaking it on a highway, and thus, to de-sterilization upon transformation
to a very fast reference frame (see Fig.~\ref{fig3}).  Also, we recall that for
Majorana neutrinos, one particular set of helicity states (right-handed neutrino
states whose antiparticles are left-handed anti-neutrinos) acquire a very large
mass within the seesaw mechanism (see p.~100 of Ref.~\cite{BaMaWh2012}), 
leading to a much more effective suppression
involving a power of $m/\Lambda_{\rm GUT}$. 

One might refer to the underlying questions, which have been discussed in
Refs.~\cite{Fe1998q76,Ho1998q76,GoGo2011}, as the ``helicity reversal paradox'',
the ``autobahn helicity paradox'', or, with reference to Ref.~\cite{LEPUS},
the ``rabbit paradox''.  One might argue that the helicity reversal
of Sec.~\ref{sec2} is described by a Lorentz transformation and does not
constitutes a paradox. The same argument, though, would otherwise apply to the
``twin paradox'' of time dilation which can be ``resolved'' by pointing to the
asymmetry of the problem, in view of the necessary acceleration of the space
craft as the ``younger'' twin reverses the course.  Also, the ``Ehrenfest
paradox'' (see Ref.~\cite{Eh1909}) which applies to the fast rotation speeds
exceeding the speed of light in revolving Lorentz frames, for points
sufficiently displaced from the rotation axis, can be resolved by carefully
analyzing the physical interpretation of the
coordinates~\cite{He2000,NZRZGh2013}.  However, it appears to be customary to
refer to an intriguing, counter-intuitive (relativistic) phenomenon as a
``paradox'', and this is why we would advocate the designation of a somewhat
paradoxical status to ``question~\#76'' (see Refs.~\cite{Fe1998q76,Ho1998q76})
connected with the helicity reversal upon a Lorentz boost.  In any case, the
question becomes a true paradox if we additionally assume that the apparent
absence of right-handed neutrinos in nature is due to some more fundamental
reason such as the exclusion from the physical spectrum by a Gupta--Bleuler
condition~\cite{JeWu2013isrn} rather than practical difficulties incurred in
the detection of the right-handed states.  In Sec.~3 of Ref.~\cite{KoMe2012},
the problem is solved by simply projecting the neutrino field, with general
Dirac and Majorana mass terms, by projecting the entire neutrino field onto its
left-handed component.  We might add that the ``almost'' sterile, light,
right-handed, overtaken Dirac neutrinos cannot be interpreted as anti-neutrinos
because the energy of the left-handed Dirac neutrinos does not change sign upon
a Lorentz boost given in Eq.~\eqref{boost}.

A superluminal neutrino, described by a generalized Dirac equation with a
pseudo-scalar mass term (see Sec.~\ref{sec3}), has the potential to address a
few physical questions connected with neutrinos. The ``wrong helicity states''
are completely suppressed from the physical neutrino fields due to
Gupta--Bleuler condition, and this remains true for infinitesimally
superluminal neutrinos with a very small tachyonic mass, because superluminal
particles always remain superluminal upon Lorentz transformation.  One cannot
overtake an ever-so-slightly superluminal neutrino.  This follows from the
Lorentz transformation, by which superluminal particles always remain
superluminal under a transformation between subluminal reference frames.  The
tachyonic Dirac equation allows for plane-wave solutions, so that essentially
nothing has to be altered in electroweak theory if we accept the fact that
subluminal and superluminal particles may couple through Lorentz-invariant
interactions.  Perhaps, one might investigate, in the future, a scenario where
neutrinos are superluminal ``but not as superluminal as recent false and
retracted experimental claims'' would otherwise suggest. 

Experimental determinations of the neutrino mass square have typically resulted
in negative expectation values, yet, compatible with zero within experimental
error bounds. Also, experiments on the neutrino propagation
velocity have typically yielded results compatible with $v=c$ within error.
Quoting Hilbert, we ``must know'' and ``will know'' as the experimental accuracy of
time-of-flight and neutrinoless beta decay experiments improves.

%
%
\section*{Acknowledgements}

The authors acknowledge helpful conversations with R. J. Hill, W.~Rodejohann,
T.~N.~Moentmann and I.~N\'{a}ndori.  Furthermore, insightful remarks from an
anonymous referee are acknowledged. This paper is in part based in part on
discussions held during the First International Conference on Logic and
Relativity, Alfr\'{e}d R\'{e}nyi Institute of Mathematics, September 2012,
Budapest, Hungary.  Thus authors wish to acknowledge insightful conversations
with Professors Istv\'{a}n Nemeti and Gergely Sz\'{e}kely.  The work was
completed during the summer months of~2013 when one of the authors (U.D.J.) was
primarily working at the ATOMKI institute of the Hungarian Academy of Sciences
in Debrecen; kind hospitality is gratefully acknowledged.  Support by the
National Science Foundation (Grant PHY--1068547) and by the National Institute
of Standards and Technology (precision measurement grant) is also gratefully
acknowledged.

\appendix 

%
%
\section{Fast Rotations and Helicity Flip}
\label{appa}

In Sec.~\ref{sec2}, we have calculated the helicity flip 
of a neutrino, overtaken on a highway. One might think that the 
same helicity flip might occur in a reference frame which 
rotates around the propagation direction (assumed to be 
the $z$ axis) of the neutrino, imagining that the 
spin of a particle can be viewed as the rotation of a spinning top.
Let us first present an intuitive, classical argument 
suggesting that a helicity flip cannot occur upon transformation 
to the rotating frame and then 
supplement this argument by a full quantum calculation, in the 
spinor formalism.

The classical argument is based on the following observation:
Imagine the neutrino as a spinning top with a (classical)
angular momentum component $L_z = \tfrac12 \, \hbar$.
If the mass of the spinning top is concentrated on a ring 
displaced by a radius $r$ from the central rotation axis,
and $m$ is the mass of the spinning classical ``neutrino'',
then the angular momentum is $L_z = r m \, v$, where
$v$ is the rotation speed of the mass distribution.
The required speed is equal to the speed of light at 
a radius governed by the equation
\begin{equation}
\label{teach}
L_z = \tfrac12 \, \hbar = r m \, v = r m \,c \,,
\quad 
r = \frac{\hbar}{2 m c} = \frac{\lambdabar}{2} \,,
\end{equation}
where $\lambdabar$ is the reduced Compton wavelength.
One cannot localize a quantum particle better than its 
Compton wavelength, and therefore the quantum-classical 
analogue of a spinning top makes sense provided we 
assume that the particle's mass distribution 
is displaced from the central rotation axis by a distance 
commensurate with the particle's Compton wavelength.
Conversely, only an ``outside'' observer could potentially 
see the neutrino spinning in the opposite direction 
and reverse its helicity; if we are ``inside'' the 
neutrino's Compton wavelength then the concept of a 
classical spinning top becomes meaningless.
Equation~\eqref{teach} teaches us that if we are to 
overtake the rotating spinning top by spinning around the axis 
of rotation faster than the neutrino spin does, and revert its 
helicity, then we have to rotate about the symmetry axis with a 
velocity  faster than the speed of light.
This is impossible for a classical observer starting from rest,
according to special relativity and therefore, 
the classical consideration strongly suggest that 
helicity reversal in a rotating frame is impossible.
(With reference to Sec.~\ref{sec3} of this article, 
we should reconfirm that, by contrast, tachyons are never at rest, and remain 
superluminal in reference frame connected by a 
proper orthochronous Lorentz 
transformation~\cite{BiDeSu1962,BiSu1969,Re2009,Bi2009,Bo2009}.)

Let us supplement this classical consideration by a full quantum 
calculation. Setting aside possible complications due to 
rotation speeds faster than $c$ in rotating frames 
(otherwise known as the Ehrenfest 
paradox~\cite{Eh1909,He2000,NZRZGh2013}), we write 
down the Lorentz and spinor transformation for the transformation 
into the rotating frame in full analogy to 
Eq.~\eqref{boost} as follows,
\begin{subequations}
\label{rotation}
\begin{align}
{\Lambda^\mu}_\nu = & \;
\left( \begin{array}{cccc}
1 & 0 & 0 & 0 \\
0 & \cos(\omega_0 \, t) & \sin(\omega_0 \, t) & 0 \\
0 & -\sin(\omega_0 \, t) & \cos(\omega_0 \, t) & 0 \\
0 & 0 & 0 & 1 \\
\end{array} 
\right) \,,
\\[0.133ex]
S(\Lambda) = & \; \left( \begin{array}{cccc}
\exp\left(\frac{\ii \, \omega_0 \, t}{2} \right) & 0 & 0 & 0 \\
0 & \exp\left(-\frac{\ii \, \omega_0 \, t}{2} \right) & 0 & 0 \\
0 & 0 & \exp\left(\frac{\ii \, \omega_0 \, t}{2} \right) & 0 \\
0 & 0 & 0 & \exp\left(-\frac{\ii \, \omega_0 \, t}{2} \right) \\
\end{array} 
\right) \,.
\end{align}
\end{subequations}
Here, $\omega_0$ is the angular frequency of the rotation.
We perform a passive rotation, i.e., the transformed 
coordinates correspond to the rotating frame, which rotates 
about the $z$ axis in the counter-clockwise 
(mathematically positive) direction.
The generators for the rotation, in the sense of 
Eq.~\eqref{boostgen}, read as
$\omega_{21} = -\omega_{12} = \omega_0 \, t = \varphi$,
with all other elements vanishing
(here, $\varphi$ is the azimuthal rotation angle).
The wave vector $\vec k = k \, \hat e_z$ is invariant 
under the rotation, 
and the effect of the rotation into the rotating frame
can be illustrated for the positive-helicity state
$u_+(k \, \hat e_z)$ defined according to Eq.~\eqref{UU1}
as follows,
\begin{align}
\label{inspection}
& u_+(k \, \hat e_z) \, 
\ee^{\ii k\, z - \ii E \, t} =
\left( \begin{array}{c}
\sqrt{\dfrac{E + m}{2 \, m}} \; 
\left( \begin{array}{c} 1 \\ 0 \end{array} \right) 
\\[2.33ex]
\sqrt{\dfrac{E - m}{2 \, m}} \;
\left( \begin{array}{c} 1 \\ 0 \end{array} \right) 
\end{array} \right) 
\nonumber\\[0.133ex]
& \; \to
S(\Lambda) \, u_+(k \, \hat e_z) \,
\ee^{\ii k\, z - \ii E \, t} =
\ee^{\ii \, \omega_0 \, t/2} \;
\left( \begin{array}{c}
\sqrt{\dfrac{E + m}{2 \, m}} \; 
\left( \begin{array}{c} 1 \\ 0 \end{array} \right) 
\\[2.33ex]
\sqrt{\dfrac{E - m}{2 \, m}} \;
\left( \begin{array}{c} 1 \\ 0 \end{array} \right) 
\end{array} \right) \,
\ee^{\ii k\, z - \ii E \, t}  \,.
\end{align}
We here indicate the dependence of the bispinor wave function
on the spatio-temporal phase factor explicitly; the 
expression $k \,z - E t$ is invariant under the rotation 
about the $z$ axis.
The energy, as measured in the rotating frame, is decreased
by an amount $-\omega_0 t /2$, by virtue of the 
definition of the Hamilton operator as $H = \ii \partial_t$.
This is in full analogy to a classical consideration, 
where the angular frequency of the spinning top (we note the
counter-clockwise classical rotation spin for positive helicity) 
is decreased in the rotating frame.
This situation is reversed for negative helicity, 
where according to Eq.~\eqref{rotation} the time-dependent phase 
factor is $\exp\left(-\ii \, \omega_0 \, t/2 \right)$, and 
the rotation speed is higher in the rotating frame,
resulting in an energy increase by $\omega_0 \, t/2$.

An inspection of the transformed wave function 
in Eq.~\eqref{inspection} shows that 
the helicity in the rotating frame is 
not reversed, irrespective of the magnitude of the 
angular rotation frequency $\omega_0$.
Namely, one can easily show that the operator that 
measures the helicity for a particle propagating in the 
positive $z$ direction is 
$\vec \Sigma \cdot (k \, \hat e_z)/k = \Sigma^3$,
where
\begin{equation}
\Sigma^i = \left( \begin{array}{cc} \sigma^i & 0 \\ 0 & \sigma^i 
\\ \end{array} \right) \,,
\quad 
\mbox{$i=1,2,3$\,,}
\end{equation}
is a $4 \times 4$ spin matrix.
This finding in turn confirms the classical argument 
regarding the rotation about the central axis faster than the 
speed of light, which is impossible for massive 
observers starting from rest.

Let us also address the question of a conceivable 
helicity reversal in the transformation into a 
frame which rotates at an angular frequency
described by the vector 
\begin{equation}
\vec \omega = \omega_0 \, \hat n \,,
\qquad 
\hat n = 
\sin\theta_\omega \, \cos \varphi_\omega \; \hat e_x + 
\sin\theta_\omega \, \sin \varphi_\omega \; \hat e_y +
\cos\theta_\omega \; \hat e_z \,.
\end{equation}
Under the rotation with angle $\vec \varphi= \vec \omega \, t
= \omega_0 \, \hat n \, t$, with Lorentz generators
\begin{equation}
\Lambda_{\mu\nu} \approx \delta_{\mu\nu} + \omega_{\mu\nu} \,,
\qquad
\omega_{\mu\nu} = 
\left( \begin{array}{cccc}
0 & 0 & 0 & 0 \\
0 & 0 & \varphi_z & -\varphi_y \\
0 & -\varphi_z & 0 & \varphi_x \\
0 & \varphi_y & -\varphi_x & 1 \\
\end{array} 
\right) \,,
\end{equation}
the initial wave vector $\vec k = k \, \hat e_z$ transforms into 
$\vec k' = k'_x \, \hat e_x + k'_y \, \hat e_y + k'_z \, \hat e_z$
with 
\begin{subequations}
\begin{align}
k'_x =& \; \sin(\theta_\omega) \, \left(2 \cos(\theta_\omega) \cos(\varphi_\omega) 
\sin^2\left(\frac12 \omega_0 t\right) 
- \sin(\varphi_\omega) \, \sin(\omega_0 t) \right) \,,
\\[0.133ex]
k'_y =& \; \sin(\theta_\omega) \, \left(2 \cos(\theta_\omega) \cos(\varphi_\omega) 
\sin^2\left(\frac12 \omega_0 t\right) 
+ \cos(\varphi_\omega) \, \sin(\omega_0 t) \right) \,,
\\[0.133ex]
k'_z =& \; \cos^2(\theta_\omega) + \cos(\omega_0\,t) \, \sin^2(\theta_\omega) \,.
\end{align}
\end{subequations}
The helicity spinor transforms as follows,
\begin{align}
\label{conclude}
a_+(k \, \hat e_z) =& \; \left( \begin{array}{c} 1 \\ 0 \end{array} \right)  
\to
\left( \cos(\tfrac12 \, |\vec \varphi|) \; \bm{1}_{2\times 2} + 
\ii \, \frac{ \sin(\tfrac12 \, |\vec \varphi|) }{ |\vec \varphi | } \,
\vec\sigma \cdot \vec \varphi \right) \,
\left( \begin{array}{c} 1 \\ 0 \end{array} \right) 
\nonumber\\[1ex]
=& \;
\left( \begin{array}{c} 
\cos(\tfrac12 \omega_0 t) + \ii \, \cos(\theta_\omega) \, 
\sin(\tfrac12 \omega_0 t) \\[0.5ex]
\ii \, \ee^{\ii \varphi_\omega} \, \sin(\theta_\omega) \, 
\sin(\tfrac12 \omega_0 t) \end{array} \right) 
= 
\ee^{\ii \, \arctan\left( \cos(\theta_\omega) \, 
\tan(\tfrac12 \, \omega_0 \, t) \right)} \, a_+( \vec k') \,.
\end{align}
The phase factor simplifies to $\exp(\ii \omega_0 \, t/2)$ for $\theta_\omega = 0$.
From Eq.~\eqref{conclude}, one might otherwise conclude that the 
lower component (negative helicity) of the spinor $a_+(\vec k)$ gets populated,
upon a Lorentz transformation involving a rotation, with the amplitude 
\begin{equation}
\left|\ii \, \ee^{\ii \varphi_\omega} \, \sin(\theta_\omega) \, 
\sin(\tfrac12 \omega_0 t) \right|^2 =
\sin^2(\theta_\omega)  \, \sin^2(\tfrac12 \omega_0 t)
\end{equation}
and interpret this as a rotation-induced helicity flip
[see Eq.~(7) of Ref.~\cite{AhLe2013}].
However, this is not the case, because one needs to interpret the transformed 
helicity spinor in terms of the transformed wave 
function [factor $\exp(\ii \, \vec k' \cdot \vec r')$], where it reproduces the 
positive-helicity spinor $a_+( \vec k')$, up to 
the phase factor $\exp[\ii \, \arctan\left( \cos(\theta_\omega) \, 
\tan(\tfrac12 \, \omega_0 \, t) \right)]$.
Finally, the bispinor transforms as follows,
\begin{align}
\label{inspection2}
& u_+(k \, \hat e_z) \, 
\ee^{\ii k\, z - \ii E \, t} =
\left( \begin{array}{c}
\sqrt{\dfrac{E + m}{2 \, m}} \; 
\left( \begin{array}{c} 1 \\ 0 \end{array} \right) 
\\[2.33ex]
\sqrt{\dfrac{E - m}{2 \, m}} \;
\left( \begin{array}{c} 1 \\ 0 \end{array} \right) 
\end{array} \right) 
\to S(\Lambda) \, u_+(k \, \hat e_z) \, 
\ee^{\ii \vec k' \cdot \vec r' - \ii E \, t} 
\nonumber\\[0.133ex]
& \; \qquad =
\ee^{\ii \, \arctan\left( \cos(\theta_\omega) \, 
\tan(\tfrac12 \, \omega_0 \, t) \right)} \;
\left( \begin{array}{c}
\sqrt{\dfrac{E + m}{2 \, m}} \; a_+(\vec k')
\\[2.33ex]
\sqrt{\dfrac{E - m}{2 \, m}} \; a_+(\vec k')
\end{array} \right) \,
\ee^{\ii \vec k'\cdot \vec r' - \ii E \, t}  \,.
\end{align}
(The energy is invariant under the rotation.)
This formula confirms once more that a rotation of the 
reference system cannot induce a helicity flip of the neutrino;
our arguments are in full agreement with 
the discussion following Eq.~(2.341) 
in Sec.~2.10.1 of Ref.~\cite{GiKi2007}
but is at variance with the 
conclusions recently reached in Ref.~\cite{AhLe2013}. 
One cannot avoid a Lorentz boost (as opposed to a rotation), as described 
in Sec.~\ref{sec2}, to induce the elusive helicity reversal 
of the Dirac bispinor.

%
%
\section{Neutrinos and Generalized Dirac Hamiltonians}
\label{appb}

Having clarified our alternative suppression mechanism 
for right-handed neutrinos, described by the tachyonic 
Dirac equation.  we should now also include a few remarks on neutrino mass mixing in the 
tachyonic sector.
The possibility of three-generation neutrino flavour-mass mixing has been indicated in 
Refs.~\cite{Po1957,MaNaSa1962},
and parameterized in the PMNS matrix.
Reviews on the theory of massive neutrinos can be found in 
Refs.~\cite{BiPe1987,BiGiGr1999,Ro2012}.
A discussion of the mixing matrix in concise form has been 
given in Ref.~\cite{Bi2006}.
One usually assumes that the neutrino flavour eigenstate $\nu_{\ell L}(x)$
with $\ell = e,\mu,\tau$
is a linear superposition of the mass eigenstates $\nu_{iL}$
with $i =1,2,3$ as follows,
\begin{equation}
\nu_{\ell L}(x) = \sum_{i=1}^3 U_{\ell i} \; \nu_{iL}(x) \,.
\end{equation}
If CPT symmetry holds and 
if there are no highly non-standard interactions affecting 
neutrinos and anti-neutrinos differently, then
the antineutrino eigenstates are mixed with the complex conjugate 
matrix,
\begin{equation}
{\overline \nu}_{\ell L}(x) = \sum_{i=1}^3 U^*_{\ell i} \; {\overline \nu}_{iL}(x) \,.
\end{equation}
The tachyonic Dirac equation is obtained from the ordinary 
Dirac equation based on the replacement $m \to \gamma^5 \, m$.
One could thus introduce flavour-mass mixing matrix 
in the tachyonic sector. In particular, the 
result~\cite{BiGiGr1999}
\begin{equation}
P(\nu_\ell \to\nu_{\ell'}) = 
\delta_{\ell \ell'} - 2 \, {\rm Re} \sum_{i>k} 
U_{\ell' i} \, 
U^*_{\ell i} \, 
U^*_{\ell' k} \, 
U_{\ell k} \, 
\left[ 1 - \exp\left(-\ii \frac{\Delta m^2_{ki} \, L}{2 \, E} \right) \right]
\end{equation}
for neutrino oscillations in a baseline experiment with length $L$,
remains valid in the tachyonic sector, provided one assumes 
a single, tachyonic mass term of the form $\gamma^5 \, m$ and 
parameterizes the mixing matrix accordingly.

With an infinitesimally superluminal neutrino, it is very hard to send
information into the past, because of the smallness of neutrino interaction
cross sections at low energy.  Colloquially speaking, the dilemma is that
high-energy tachyonic neutrinos approach the light cone and travel only
infinitesimally faster than light itself ($E = \sqrt{\vec p^2 c^2 + m^2 c^4}
\approx |\vec p|\, c$ for high energy).  Their interaction cross sections may
be sufficiently large to allow for good detection efficiency but this is
achieved at the cost of sacrificing the speed advantage. Low-energy tachyonic
neutrinos may a substantially faster than light but their interaction cross
sections are small and the information sent via them may be  lost.  A more
quantitative argument is given in Ref.~\cite{JeEtAl2014}.  If we raise the
impossibility to send information into the past to a postulate, then tachyonic
spin-$1/2$ particles, if they exist, have to be very light. 

According to the Feynman prescription, the propagator of the quantized Dirac
field has an advanced, strictly speaking {\em acausal} part which describes
antiparticles moving backward in time, and it also has a causal, retarded part,
describing particles moving forward in time. However, the only physically
relevant amplitudes predicted by theory are elements of the scattering matrix
($S$-matrix).  Antiparticles moving backward in time are reinterpreted as
entities moving forward in time, with all kinetic variables (energy, momentum)
changing sign~\cite{Fe1949ee,St1942}.  This is a manifestation of the principle
of reinterpretation. The transition amplitude predicted by $S$ matrix theory
connects two space-time points, and the time coordinate of one of the events
happens to be earlier than the other.  Particles, seen by an observer, always
move forward in time because it is always possible to identify the ``earlier''
event whose time coordinate is less than that of the other event. If the
particles described by the theory are subluminal (move slower than light), then
this reinterpretation is undisputed within the community and forms one the core
foundations of experimentally verified quantum field theory.  This concept has
been generalized to superluminal particles~\cite{BiSu1969}.

The compatibility of faster-than-light travel with the axioms of special
relativity has been discussed in
Refs.~\cite{BiDeSu1962,BiSu1969,Re2009,Bi2009,Bo2009,Sz2012,MaStSz2014}.  Via a
geometric construction (Minkowski diagram), one can show that a superluminal
velocity remains superluminal if one changes Lorentz frames.  In particular,
the Einstein velocity addition theorem remains valid in the superluminal
world~\cite{BiDeSu1962,BiSu1969}.  Nimtz and
coworkers~\cite{EnNi1992,NiSt2008,Ni2009} claim to have demonstrated in their
(disputed) experiments that electromagnetic signal propagation with up to four
times of the speed of light is possible if one is willing to accept exponential
damping of the signal (tunneling effect), i.e., over short distances.
 
The quantum field theory of superluminal particles is plagued with a series of
problems, and it was soon realized that not all of the so-called
Osterwalder--Schrader axioms~\cite{OsSc1973} can be retained if one tries to
incorporate tachyonic particles into field theory.  In a series of recent
papers~\cite{JeWu2012epjc,JeWu2012jpa,JeWu2013isrn}, we have used the concept
of a Lorentz non-invariant vacuum state, breaking one of the
Osterwalder--Schrader~\cite{OsSc1973}  axioms, but we have retained absolute
conformity with Lorentz covariance and the special theory of relativity.  Also,
we have offered an alternative picture (``re-reinterpretation'') in Sec.~4 of
Ref.~\cite{JeWu2012epjc}) where we work with a Lorentz-invariant vacuum while
transforming only the space-time argument of the creation and annihilation
operators, but not, as it would seem necessary otherwise, some of the
annihilation operators of the tachyonic field into creation operators and vice
versa.  The Hamiltonians describing faster-than-light particles are
pseudo-Hermitian.  The concept of pseudo--Hermiticity was introduced in
Ref.~\cite{Pa1943}.  Coincidentally, the only particle which is a candidate for
superluminality, the neutrino, has been observed in only one helicity
(left-handed), while antineutrinos come in right-handed helicity. This is
precisely what our anticommutator relation~\eqref{anticom} predicts, and
postulating this anticommutator is the only possibility we found in order to
complete the spinor sum in Eq.~\eqref{r2}, where the prefactor $(-\sigma)$ is
necessary in order to obtain the positive-energy and negative-energy
projectors.  No further alterations are necessary for the theory of weak
interactions because Lorentz invariance is fully conserved.  In principle, the
numerical value of the tachyonic mass term could be determined using more
precise experiments on neutrino flight times or by looking at the end point of
the tritium beta decay experiments more accurately.

If we assume that if the neutrino is superluminal, then the neutrino is not
equal to its own antiparticle, and neutrinoless double beta decay is forbidden.
While this conclusion somewhat depends on the precise equation proposed for the
description of the neutrino, none of the currently proposed
equations~\cite{ChHaKo1985,Ch2000,Ch2002,JeWu2012epjc,JeWu2012jpa,JeWu2013isrn}
is charge conjugation invariant, so that there are no
charge-conjugation-invariant Majorana solutions of the tachyonic spin-$1/2$
equations.  We recall that experimental evidence for the observation of
neutrinoless double beta
decay~\cite{Fi1949,FiSc1952,KKEtAl2004,KKKr2006,KKDiHaKr2001,FeStVi2002,ZdDaTr2002}
is disputed, and recent measurements set even stricter bounds for the magnitude
of the Majorana mass terms (for an overview see pp.~176 ff.~of Ref.~\cite{Zu2012}). 
Relatively recent results include data from Cuoricino (see Ref.~\cite{Gi2012}),
from the EXO-200 collaboration~\cite{AuEtAl2013},
from KamLAND-Zen~\cite{GaEtAl2013} and GERDA~\cite{MaEtAl2013}.

%
%
\section{Majorana Equation}
\label{appc}

Let $\omega(x)$ denote a two-component spinor amplitude.
In the conventions of Ref.~\cite{Pa2011ajp},
where the chiral representation of the Dirac algebra is employed,
the Majorana equation reads [see Eq.~(6.25) of Ref.~\cite{Pa2011ajp}]
\begin{equation}
\label{onlytwo}
\overline{\sigma}^\mu \, \partial_\mu \omega(x) + m \, \sigma^2 \, \omega^*(x) = 0 \,.
\end{equation}
Here, $\overline{\sigma}^\mu = (1, -\vec \sigma)$,
and the $y$ component of the $\vec \sigma$ vector is denoted as $\sigma^2$.
This is a two-component equation which clearly cannot 
be solved by a plane-wave ansatz because the first 
term would involve a factor $\exp(-\ii \, k \cdot x)$, 
whereas the second one, which involves the complex
conjugate $\omega^*(x)$, would go as $\exp(\ii \, k \cdot x)$.
Consequently, in the field operator of the Majorana particle
given in Eq.~\eqref{fk}, the
plane-wave spinor wave functions multiplying the creation and annihilation
operators in the Fourier decomposition 
are not plane-wave eigenstates of the Majorana
equation. 

In Ref.~\cite{Ka2009}, it is stated that 
the mass eigenstates of the Majorana neutrino are
not plane waves but are of the form $\nu_L + \nu_L^c$,
i.e., superpositions of left- and right-handed
states [specifically, 
see Eq.~(3) of Ref.~\cite{Ka2009}]. This is permissible if the neutrino
is its own antiparticle. 
In this context, we also recall that the charge-conjugation
invariant solutions of the massless Dirac equation,
\begin{equation}
\label{CCinv}
\Psi_+(x) = u_{+} (\vec k) \; \ee^{- \ii k \cdot x} +
v_-(\vec k) \; \ee^{\ii k \cdot x} \,, 
\quad
C \overline \Psi^{\rm T}_+(x) = \Psi_+(x) \,,
\end{equation}
and
\begin{equation}
\Psi_-(x) = u_{-} (\vec k) \; \ee^{- \ii k \cdot x} +
v_+(\vec k) \; \ee^{\ii k \cdot x} \,,
\quad
C \overline \Psi^{\rm T}_-(x) = \Psi_-(x) \,,
\end{equation}
are not plane waves with a definite four-momentum,
but superpositions of plane waves which travel 
in opposite directions. 
Charge conjugation invariance holds 
because of the relation $C \, \overline u_\pm(\vec k)  = v_\mp$
with $C = \ii \, \gamma^2 \, \gamma^0$
(in the standard representation).
So, an outgoing Majorana neutrino in a scattering process,
described by a wave with a definite four-momentum,
momentum cannot simultaneously be in an energy eigenstate of the Majorana equation.
By contrast, all eigenspinors of the Dirac
equation (with scalar or pseudo-scalar mass
terms) simultaneously constitute energy and momentum eigenstates of the 
Dirac Hamiltonian, in full compatibility with 
unperturbed electrons, positrons, heavy leptons, 
photons, heavy gauge bosons, and quarks, which are all described 
by plane-wave states proportional to $\ee^{-\ii k \cdot x}$.

\end{document}